\documentclass{article}

\usepackage{arxiv}
\usepackage[authoryear]{natbib}
\usepackage[utf8]{inputenc} 
\usepackage[T1]{fontenc}    
\usepackage{hyperref}       
\usepackage{url}            
\usepackage{booktabs}       
\usepackage{amsfonts}       
\usepackage{nicefrac}       
\usepackage{microtype}      
\usepackage{lipsum}
\usepackage{subfig}
\usepackage{float}
\usepackage{graphicx}
\graphicspath{{Pictures/}}

\title{Efficient Indexing of Meta-Data \\(extracted from educational videos)}

\author{
 Shalika Kumbham \\
  Indian Institute of Technology, Kharagpur\\
  \texttt{shalika@iitkgp.ac.in} \\
   \And
  Abhijit Debnath \\
  Indian Institute of Technology, Kharagpur\\
  \texttt{abhijitdebnath@iitkgp.ac.in} \\
  \And
 Krothapalli Sreenivasa Rao \\
  Indian Institute of Technology, Kharagpur\\
  \texttt{ksrao@cse.iitkgp.ac.in} \\
}

\begin{document}
\maketitle
\begin{abstract}
Video lectures are becoming more popular and in demand as online classroom teaching is becoming more prevalent. Massive Open Online Courses (MOOCs), such as NPTEL, have been creating high-quality educational content that is freely accessible to students online. A large number of colleges across the country are now using NPTEL videos in their classrooms. So more video lectures are being recorded, maintained, and uploaded. These videos generally contain information about that video before the lecture begins. We generally observe that these educational videos have metadata containing five to six attributes: Institute Name, Publisher Name, Department Name, Professor Name, Subject Name, and Topic Name. It would be easy to maintain these videos if we could organize them according to their categories.  The indexing of these videos based on this information is beneficial for students all around the world to efficiently utilise these videos. In this project, we are trying to get the metadata information mentioned above from the video lectures. 
\end{abstract}


\section{Introduction}
Students are going through numerous changes as they adjust to the COVID situation. The mode of classes is one of the most affected. 
A video is a digitally recorded and saved programme, movie, or other visual media product that contains moving visuals with or without audio. The frame is a single image in a sequence of pictures. In most cases, one second of video is made up of 24 or 30 frames per second (FPS). The frame is made up of the image and the time the image was exposed to the view when it was taken. The animation is created by extracting a series of frames in a row.
Creating and managing web videos is a difficult task. It would be a lot easier if we could automatically categorise them without much human interaction. It saves a significant amount of time and work. As a result, we're attempting to create an interface for obtaining the necessary data and organising it into the appropriate categories. If you give it a video, it will provide you with information from those categories.
For example, if there is an NPTEL video on Image Processing taught by Prof. K S Rao from IIT Kharagpur's Computer Science Department, we must map the information to the appropriate column, such as Publisher Name - NPTEL, Professor Name - K S Rao, Department Name - Computer Science, and so on.

\section{Literature Review}
\citet{deepkeyframehuman} talked about key frame extraction from videos (human action videos) using a deep learning approach. An algorithm has been proposed that is mostly focused on the fusion of deep features and histograms to reduce the number of redundant frames.

Because of the combined elements of human position in movement and the background, detecting representative frames in movies based on human actions is fairly difficult. \citet{deepkey} address the issue and define key frame detection as the process of identifying the video frames that contribute the most to distinguishing the underlying action category from all others. They tested their method on the UCF101 dataset, which is a difficult human action dataset, and found that it can detect critical frames with high accuracy.

\citet{optical} discusses the need to index the images in a computer system for easy access. They mention the use of OCR to achieve the same. They went over the pre-processing, segmentation, normalisation, feature extraction, classification, and post-processing aspects of an OCR system and the challenges involved

\citet{ocr} worked on improving the recognition ability of text in a video based on OCR. Machine Learning techniques were used to improve the accuracy of the recognition with suitable preprocessing.

\citet{indexing} discussed the need for video indexing and offered an approach for automatic lecture video indexing based on video OCR technology. They developed a new technique for analysing and extracting slide structure based on the geometrical information of recognised text lines.

\citet{autoindex} implemented a new video segmenter for automatic slide video structure analysis and a word detector based on weighted DCT (discrete cosines transformation).

The TIB AV-Portal (\citet{tib}) is a portal for scientific videos with a focus on technology/engineering as well as architecture, chemistry, computer science, mathematics, and physics. Other scientific disciplines are also included in the AV-Portal. The videos include computer visualizations, learning material, simulations, experiments, interviews, video abstracts, lecture, and conference recordings, and (open) audiovisual learning and teaching materials. The portal uses various automatic video analyses. The added value: these video analyses enable pinpoint searches within the video content. \\
NDLI (\citet{ndli}) is an integration platform that sources metadata from a massive number of sources to provide single window access to a wide variety of digital learning resources. Metadata summarizes basic information of a resource, which can make finding and working with particular resources easier. Sources have different standards for their metadata with a variety of resources.NDLI metadata standard has been developed to draw up this uniform schema. It is an envelope of several well-established global standards. \\
The Dublin Core Metadata Initiative (\citet{dcmi}), which formulates the Dublin Core, is a project of the Association for Information Science and Technology (ASIS\&T). ” Dublin Core” is also used as an adjective for Dublin Core metadata. It is a style of metadata that draws on multiple Resource Description Framework (RDF) vocabularies, packaged and constrained in Dublin Core application profiles.
The Dublin Core Metadata Initiative (DCMI) provides an open forum for developing interoperable online metadata standards for a broad range of purposes and business models. DCMI’s activities include
\begin{itemize}
    \item consensus-driven working groups,
\item global conferences and workshops,
\item standards liaison, and
\item educational efforts to promote widespread acceptance of metadata standards and practices.
\end{itemize}
\subsection{Existing Methods}
A similar type of problem is solved in a research article by Haojin Yang, Maria Siebert, Patrick Lühne, Harald Sack, and Christoph Meinel. It employs the keyframe detection technique, text localization, text reading, and text classification into title, subtitle/key point, normal content, and footline based on the stroke width and height of the text identified. This article is the most closely related to the issue statement we're working on. We discovered few to no works that matched our problem description.

\section{Our Methodology} 
\subsection{Objective}
To develop a dataset from NPTEL lecture videos and build a better model to categorize the videos based on variables such as Institute Name, Publisher Name, Professor Name, and Department Name. An example slide containing the four attributes is shown in Figure \ref{fig:all_attributes}
\subsection{Dataset Preparation}
We presume that the features we require are present before the introductory section, i.e., before the professor begins teaching in the lecture video. 
As a result, we limited the length of the videos to the introduction part.
For better outcomes, we meticulously collected the videos to ensure that they were not concurrent with one another. 
Furthermore, the occurrences to be categorized are entirely independent of one another.

\begin{figure}[H]
    \centering
    \subfloat[]{\includegraphics[width=0.35\textwidth]{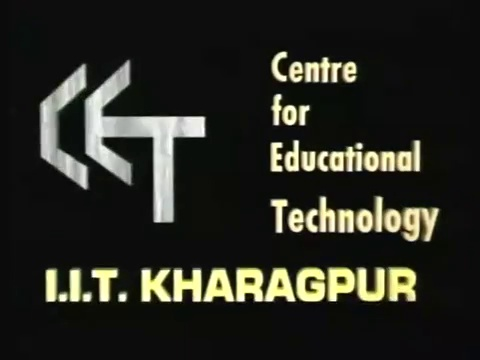}}
    \hspace{1pt}
    \subfloat[]{\includegraphics[width=0.35\textwidth]{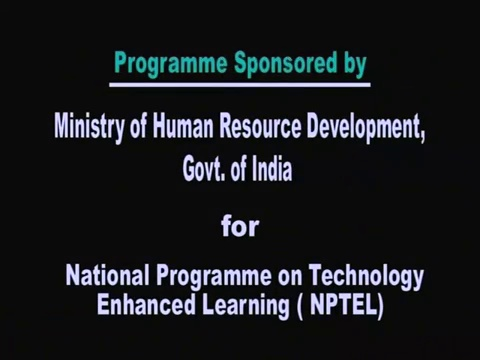}} 
    \hspace{1pt}
    \subfloat[]{\includegraphics[width=0.35\textwidth]{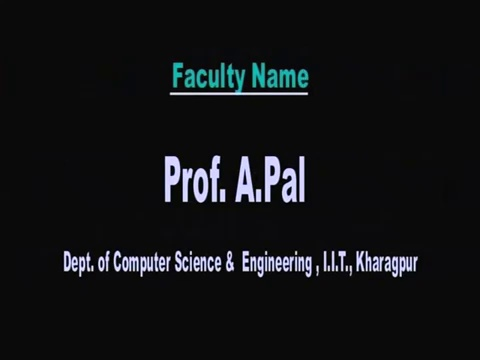}}
    \caption{(a) Frame containing Institute Name (b) Frame containing Educational Video name (c) Frame containing Professor and Department Name}
    \label{fig:different_attributes}
\end{figure}

\begin{figure}[H]
    \centering
    \includegraphics[scale=0.25]{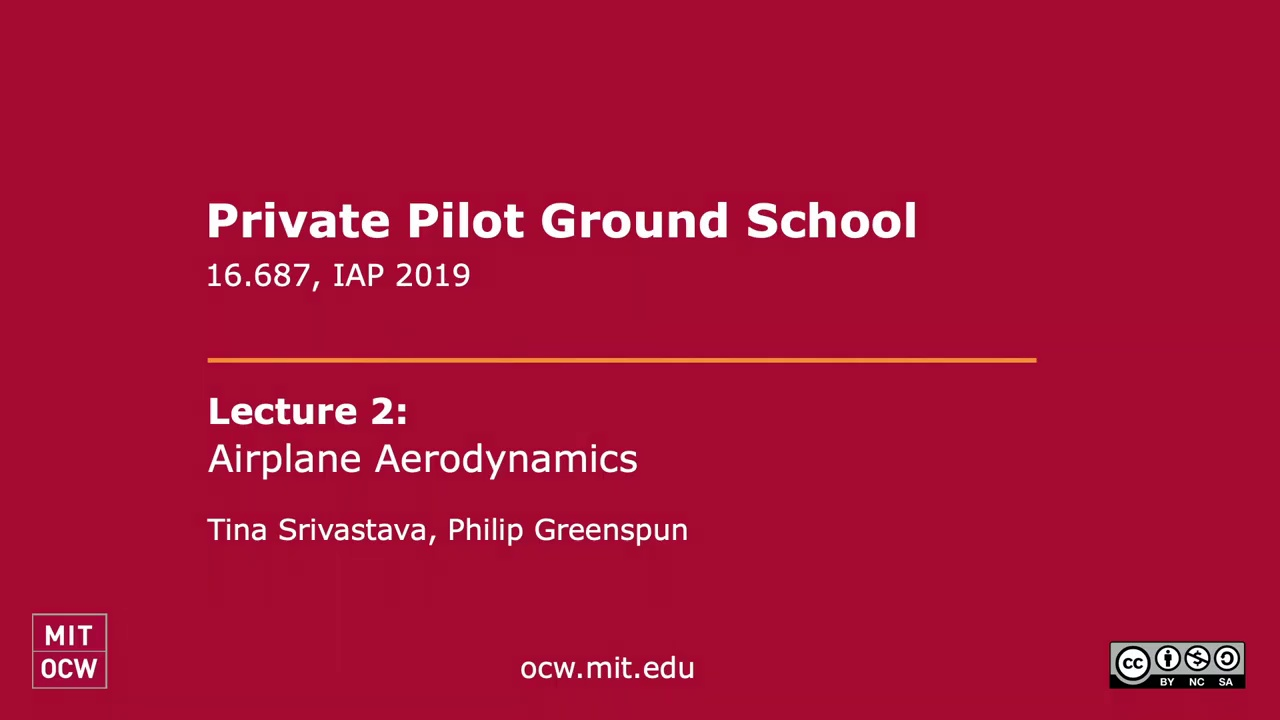}
    \caption{Image to depict missing attributes}
    \label{fig:missing}
\end{figure}

The video lecture can be categorized into 2 types:
\begin{itemize}
\item All attributes in same slide (Figure \ref{fig:all_attributes})
\begin{figure}[h!]
    \centering
    \includegraphics[width=0.6\textwidth]{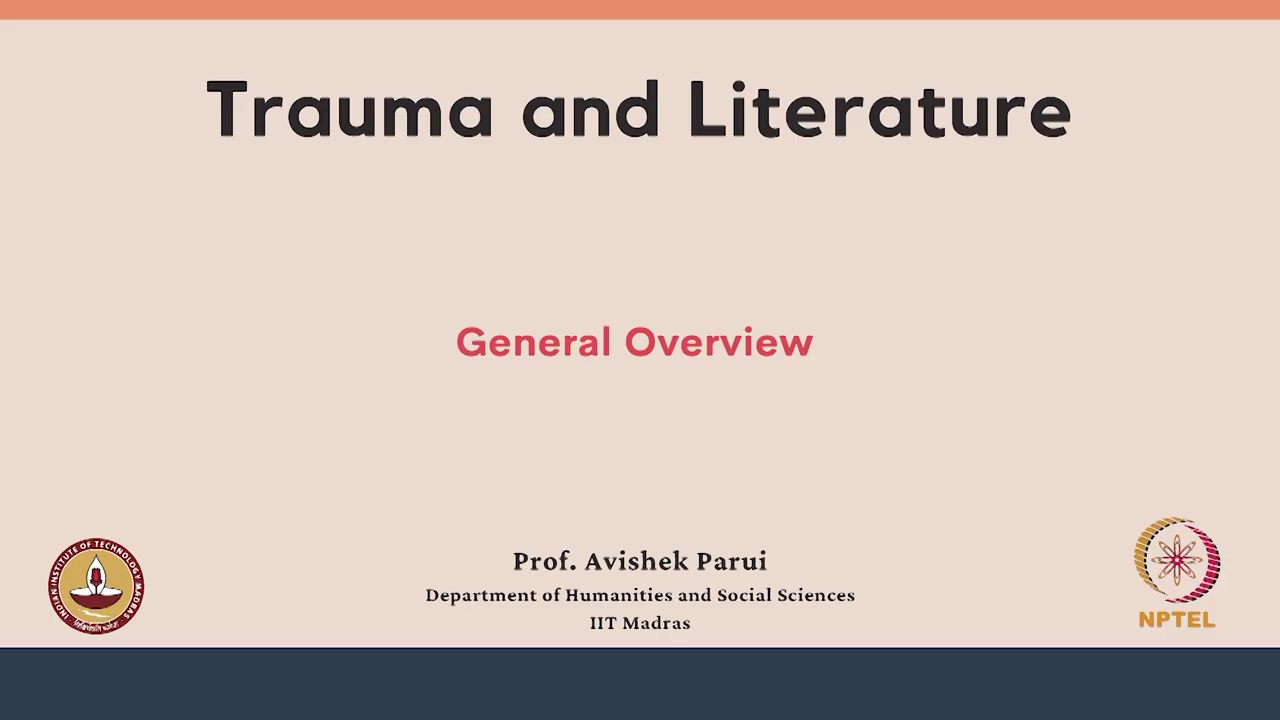}
    \caption{All the attributes in a frame}
    \label{fig:all_attributes}
\end{figure}
\item Attributes in different slides (Figure \ref{fig:different_attributes})
\end{itemize}
\vspace{10em}
All four attributes are not required to be present as seen in Figure \ref{fig:missing} nevertheless, one or more attributes may be lacking in the videos. Even with the lacking metadata, we proceed. In the case of missing metadata, we strive to find as many attributes as feasible.
\subsubsection{YouTube-dl}
YouTube-dl is an open-source video and audio download manager for YouTube and over 1000 other video hosting websites. This library is used to obtain videos, which are subsequently cropped with FFMPEG.
Steps followed:
\begin{itemize}
\item A CSV sheet containing the \emph{URL}, \emph{start-time} and \emph{end-time} is prepared.
\item \textbf{\emph{ydl.download([URL])}} is used to download the entire YouTube video.
\item Tags are then added to individual videos.
Ex:- \emph{"Id of the video, title of the video"}
\item The videos are cropped using FFMPEG and stored locally.
\newline  \emph{ffmpeg inputFile -ss startTime -to endTime -c copy outputPath/fileName-TRIM.fileExtension}
\end{itemize}

\subsection{Key-Frame Detection}
The process of extracting a frame or combination of frames that have a good representation of a shot is known as key frame extraction. It must retain the most important aspect of the shot while removing the majority of the repeated frames.
Video is a vast data object, and its analysis is critical in many domains, including video summarization, video browsing, compression, marketing, etc. We extract crucial frames from videos to decrease the dataset while retaining all important information. These extracted frames provide the most accurate and concise description of the video's content. Various key-frame extraction methods are available. We describe some of the methods below which have been experimented with.

\subsubsection{Pixel Subtraction}
Image subtraction, also known as pixel subtraction, is the process of subtracting the digital numeric value of one pixel or entire image from another image. This is done largely for one of two reasons: leveling unequal areas of an image, such as half an image with a shadow on it, or identifying differences between two photos. This change detection can be used to determine whether or not something in the image has moved. The frames are considered unique if the mean value of all pixel deviations exceeds a specific threshold.

\subsubsection{Frame count method}
OpenCV's VideoCapture function is used to capture frames from videos.
The frame count function uses the frames captured by OpenCV's VideoCapture method and saves a frame every x seconds. These frames are regarded as one-of-a-kind and are preserved for further processing. Here, x is a variable that can be modified to meet our standards and needs. However, this strategy is inefficient. Although we do not lose much information, we do receive a lot of repeated frames. As a result, it does not always decrease the data and defeats the goal of extracting keyframes.
\subsubsection{ffprobe tool}
ffprobe is a toolbox in the ffmpeg library. ffprobe collects information from multimedia streams and prints it in a human- and machine-readable format. When a video is fed into ffprobe for keyframe extraction, the frames are classified into three types: I, B, and P.

\begin{itemize}
\item Interconnected frames/ Key frames (I Frames)
\item Predictive/Predicted frames (P Frames)
\item Bi-Directional Predicted frames (B Frames)
\end{itemize}

So selecting I frames and saving them provides us with the unique frames we desire.

\subsubsection{ Mittal (2020) }
\citet{key_github} (The GitHub repository by Varun Mittal ) has a pipeline for extracting keyframes. To extract the essential frames, the pipeline goes through three steps. Each phase focuses on a specific objective in finding keyframes and restricts the number of frames to be studied in order to save time. The figure \ref{fig:Opensource} below depicts these steps, which are followed by a brief description of each phase. 

Steps followed:
\begin{itemize}
\item Step 1: Extract candidate frames
This process entails finding potential candidate frames that could become keyframes. This is done to reduce the number of frames to be processed in the next phase while retaining all relevant frames. In this stage, consecutive frame differences are calculated, followed by the frame with the greatest difference in a window of frames.
\item Step 2: Cluster similar candidate frames
Clusters of similar frames are formed in this step. Similar frames that are relatively close to one other are clustered together to form a single cluster. Before clustering, each frame is treated to extract the relevant information from it - by scaling the frame, changing it to greyscale, and with some other preprocessing. All that is done to obtain the most useful or significant information from the frame.
\item Step 3: Choose the best frames from each cluster.
Finally, the best frame from each cluster, as well as all frames that could not be clustered, are identified as keyframes in this step. The best frame is chosen based on the brightness and image blur index. Because all frames in a cluster have similar content, all other frames except the best frame from each cluster are discarded.
\end{itemize}

For all the attributes - Institute Name, Publisher Name, and Department Name - the efficiency of the FFprobe tool is higher than the \citet{key_github} 
\begin{figure}[H]
    \centering
    \includegraphics[scale=0.3]{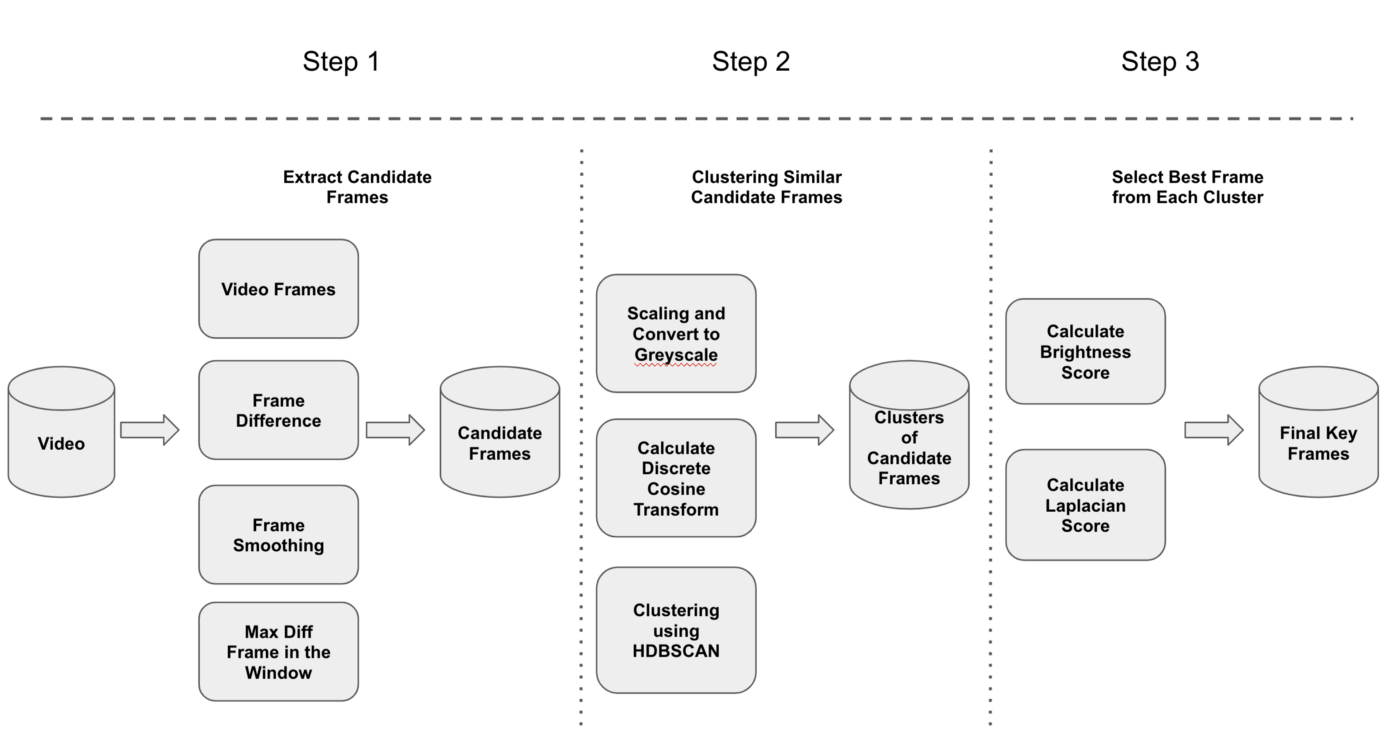}
    \caption{Steps for keyframe detection}
    \label{fig:Opensource}
\end{figure}

\subsection{Text Detection}
The process of translating text in an image to string representation is known as text detection. We employ Optical Character Recognition, often known as OCR, for this. We experimented with two types of OCRs.

\begin{itemize}
    \item Easy OCR
    \item Tesseract OCR
\end{itemize}

\subsubsection{Easy OCR}
EasyOCR, as the name implies, is a Python library that enables computer vision engineers to perform Optical Character Recognition with ease.
When it comes to OCR, EasyOCR is by far the easiest approach to implementing Optical Character Recognition.
EasyOCR is a lightweight model that performs well on photos with a slightly lower resolution.

\subsubsection{Tesseract OCR}
Tesseract – the most popular and high-quality OCR library, is an optical character recognition engine with open-source code.
Tesseract OCR employs artificial intelligence for text search and image identification.
In order to recognize words, the Tesseract OCR engine employs language-specific training data. Like the human brain, the OCR algorithms favor words and sentences that frequently appear together in a given language, and they perform well on high-resolution photos.

We used various pre-processing procedures before sending them to both OCRs, such as
\begin{itemize}
\item Converting to binary image
\item Blurring the images with Gaussian filter
\item Detecting edges through different models
\end{itemize}
We discovered that preprocessing had little effect on enhancing accuracy because the image quality was poor.
We also discovered that Easy OCR produces better results for low-quality images. We can deduce from this that Easy OCR performs some in-built pre-processing to improve the readability of the text.
\subsection{Indexing System}
A dictionary-based technique was utilized to categorize three qualities, Publisher Names, Institute Names, and Department Names. For these three qualities, a library called fuzzy-wuzzy is used.
\subsubsection{fuzzy-wuzzy}
Fuzzy String Matching, also known as Approximate String Matching, is the process of locating strings that match a pattern as closely as possible. The technique has a wide range of applications, including spell-checking, DNA analysis and detection, spam detection, plagiarism detection, etc.
Fuzzy-wuzzy is a Python module that calculates the differences between sequences and patterns using Levenshtein Distance.

In this case, fuzzy-wuzzy is utilized to match and map the words in the categories mentioned above in the text produced by EasyOCR.
We apply a predetermined set of rules and the Spacy NER Model to map the fourth attribute, Professor Name.
Some general assumptions for developing the set of rules are as follows:
\begin{itemize}
\item Professor Name is almost always written in a single line
\item Professor Name has either Prof. or Dr. as a prefix
\item Professor Name always comes in the line succeeding the word "by".
\end{itemize}

\subsubsection{Spacy NER model}
Named Entity Recognition (NER) is an application of Natural Language Processing (NLP) that processes and understands enormous amounts of unstructured human language. Entity identification, entity chunking, and entity extraction are other terms for the same thing.
SpaCy is a highly efficient statistical method for NER in Python that can assign labels to consecutive sets of tokens. It has a default model that can recognise a wide range of named or numerical things such as person, organisation, language, event, and so on.


So we used the Spacy NER model to find professors' name where the set of rules do not return the professor's name. It works well with English-origin names so works well with MIT videos. It is also used in the videos where the set of rules/assumptions do not work, to atleast retrieve the professor's name up to some extent.  We attempted to efficiently map the fourth feature, the Professor name, using the criteria and the NER model.
\section{Results} 
The table given below depicts the accuracy in the percentage of each attribute experimented with ocr(EasyOCR/Tesseract OCR), keyframe detection tool(ffprobe,\cite{key_github})
\begin{center}
\begin{tabular}{|c|c|c|c|} 
 \hline
 Category & EasyOCR , ffprobe & TesseractOCR, ffprobe & EasyOCR , \citet{key_github} \\ [0.35ex] 
 \hline\hline
 Publisher Name & \textbf{88.03} & 63.01 & 71.24 \\ 
 \hline
 Institute Name & \textbf{88.88} & 68.54 & 74.15  \\
 \hline
 Department Name & \textbf{82.47} & 65.87 & 75.07 \\
 \hline
 Professor Name & \textbf{85.89} & 64.44 & 75.60  \\ [1ex] 
 \hline
\end{tabular}
\label{tab:results}
\end{center}
From the table, we can observe that Easy OCR, ffprobe which is the baseline model is giving the best results when compared with the other two.
\bibliography{references}
\bibliographystyle{apalike}
\end{document}